\documentclass[twocolumn,superscript, address, showpacs]{revtex4}%
\usepackage{amsmath}
\usepackage{amssymb}
\usepackage{array}
\usepackage{graphicx}
\usepackage{dcolumn}
\usepackage{bm}
\usepackage{amsfonts}
\usepackage{color}
\renewcommand\arraystretch{1.5}
\usepackage[colorlinks, linkcolor=blue, anchorcolor=blue,citecolor=blue]{hyperref}
\providecommand{\U}[1]{\protect\rule{.1in}{.1in}}

\begin{document}

\title{Light transfer transitions beyond higher-order exceptional points in parity-time and anti-parity-time symmetric waveguide arrays}

\author{Chuanxun Du$^{1}$, Gang Wang$^{2}$, Yan Zhang$^{1,\dagger}$, and Jin-Hui Wu$^{1,\ast}$}
\affiliation{$^{1}$ School of Physics and Center for Quantum Sciences, Northeast Normal University, Changchun 130024, China}
\affiliation{$^{2}$ School of Physical Science and Technology, Soochow University, Suzhou 215006, China}
\affiliation{$^{\dagger}$ zhangy345@nenu.edu.cn}
\affiliation{$^{\ast}$ jhwu@nenu.edu.cn}

\begin{abstract}
We propose two non-Hermitian arrays consisting of $N=2l+1$ waveguides and exhibiting parity-time ($\mathcal{PT}$) or anti-$\mathcal{PT}$ symmetry for
investigating light transfer dynamics based on $N$th-order exceptional points (EPs). 
The $\mathcal{PT}$-symmetric array supports two $N$th-order EPs separating an unbroken and a broken phase with real and imaginary eignvalues, respectively. 
Light transfer dynamics in this array exhibits radically different behaviors, \textit{i.e.} a unidirectional oscillation behavior in the unbroken phase, an edge-towards localization behavior in the broken phase, and a center-towards localization behavior just at $N$th-order EPs.
The anti-$\mathcal{PT}$-symmetric array supports also two $N$th-order EPs separating an unbroken and a broken phase, which refer however to imaginary and real eigenvalues, respectively. Accordingly, light transfer dynamics in this array exhibits a center-towards localization behavior in the unbroken phase and an origin-centered oscillation behavior in the broken phase. 
These nontrivial light transfer behaviors and their controlled transitions are not viable for otherwise split lower-order EPs and depend on the underlying $SU(2)$ symmetry of spin-$l$ matrices.
\end{abstract}
\maketitle

\section{Introduction}

Hamiltonians in quantum mechanics are usually Hermitian and
exhibit real eigenvalues, hence convenient for describing closed
(conserved) systems with neither loss nor gain. Hermitian
Hamiltonians in combination with perturbation methods may also be
extended to describe open (non-conserved) systems involving loss or
gain; nevertheless inevitably bring extensive calculations. By
comparison, non-Hermitian Hamiltonians not only provide a
straightforward framework for describing open
systems~\cite{nonHer1,nonHer2,nonHer3,nonHer4,nonHer5}, but also are explored to reveal
quite a few unusual features unavailable for Hermitian
Hamiltonians~\cite{nonHerF1,nonHerF2,nonHerF3,nonHerF4,nonHerF5}. One prominent example is
that two or more eigenvalues and eigenvectors of a non-Hermitian
Hamiltonian can become degenerate under appropriate conditions,
yielding the so-called exceptional points
(EPs)~\cite{EP1,EP3,EP4}. So far realization and exploitation of
EPs have been well investigated in various non-Hermitian systems
especially those exhibiting, \textit{e.g.} parity-time ($\mathcal{PT}$)
symmetry~\cite{PT1,PT2,PT3,PT4} or anti-parity-time (anti-$\mathcal{PT}$)
symmetry~\cite{antiPT1,antiPT2,antiPT3}. Typically, drastic changes of
relevant phenomena would occur around EPs so that it is viable to
realize switching of optical transmission~\cite{nonHerF3,swtich1},  topological phase
transition~\cite{topo1,topo2,topo3}, ultra-sensitive
measurement~\cite{measure1,measure3,measure4}, nonreciprocal light
propagation~\cite{nonre1,nonre2,nonre3,nonre4}, etc. 
These nontrivial findings are closely associated with the orders of EPs~\cite{hEP2,hEP3,hEP4,hEP5,hEP6} because higher-order EPs usually involve more complicated physics and bring more interesting results~\cite{nonHer5,hEP11,hEP13,hEP14,hEP15,hEP16}, though also more difficult to realize owing to higher requirements on the Hamiltonian symmetry~\cite{LorzG,Haml1,Haml2}.

Hinging on the mathematical isomorphism between the Schr\"{o}dinger equation and the Maxwell paraxial wave equation, investigations on non-Hermitian physics have been extended to and are flourishing in optics and photonics in view of the great convenience on experimental realizations~\cite{nonHopt1,nonHopt2,nonHopt3,nonHopt4,nonHopt5,nonHopt6}. 
Typical optical platforms for investigating non-Hermitian phenomena include arrays of resonators and waveguides described by engineered mode-coupling matrices with complex diagonal or off-diagonal elements~\cite{array1,array2,array3}. 
It is found, for instance, light transport in two-mode $\mathcal{PT}$-symmetric waveguides with balanced gain and loss changes drastically from a reciprocal to an nonreciprocal behavior across the $2$nd-order EPs of mode-coupling matrices~\cite{nonHopt1,nonHopt2}. 
More interesting non-Hermitian characteristics and applications, such as richer topological photonics\cite{hEPtopo1}, enhanced spontaneous emission\cite{hEPemis1}, improved sensitivity of spectra and dynamics\cite{hEPsen1,hEPsen2,hEPsen3}, and
delayed sudden death of entanglement~\cite{entang1}, have also been found around higher-order EPs of mode-coupling matrices, \textit{e.g.} in $\mathcal{PT}$-symmetric waveguide arrays discussed in terms of underlying $SU(2)$ symmetry~\cite{LorzG}. Meanwhile, passive optical systems exhibiting anti-$\mathcal{PT}$ symmetry have spurred a rapidly growing interest in that they hold unique non-Hermitian features~\cite{antiPTF1,antiPTF2,antiPTF3,antiPTF4} yet without the need of gain~\cite{loss2,loss3,loss4} limiting realistic implementations and applications. 
In this regard, multi-mode anti-$\mathcal{PT}$-symmetric
waveguides deserve further studies on higher-orders EPs and
nontrivial light transport.

In this work we consider two non-Hermitian arrays consisting of $N=2l+1$ waveguides in the regimes of $\mathcal{PT}$ symmetry and anti-$\mathcal{PT}$ symmetry, respectively, for achieving two highest $N$th-order EPs and examining a few nontrivial light transfer behaviors.
Mode-coupling matrices with imaginary diagonal (sub-diagonal) elements are constructed here for describing the $\mathcal{PT}$-symmetric (anti-$\mathcal{PT}$-symmetric) waveguide array on the basis of two
spin-$l$ matrices~\cite{Haml1,Haml2} embodying the underlying $SU(2)$ symmetry. 
Different from the ballistic (breathing-like) light
transport in a Hermitian waveguide array, the light transport in our $\mathcal{PT}$-symmetric waveguide array can be switched between a unidirectional oscillation behavior in the unbroken phase and an edge-towards localization behavior (related to the skin effect~\cite{skin1}) in the broken phase through a center-towards localization behavior at one $N$th-order EP. 
Such a sensitive transition of light transfer behaviors becomes impossible when both $N$th-order EPs are split into a series of lower-order EPs by modifying somehow the corresponding mode-coupling matrix. 
Compared to the $\mathcal{PT}$-symmetric waveguide array with balanced gain and loss, the anti-$\mathcal{PT}$-symmetric waveguide array with passive couplings~\cite{loss3,loss5,loss6,loss7} supports another transition between distinct light transfer behaviors around one $N$th-order EP: a center-towards localization behavior even in the unbroken phase and an origin-centered oscillation behavior in the broken phase.

\section{$\mathcal{PT}$-symmetric waveguide array}

\begin{figure}[tbp]
\centering\includegraphics[width=8.5 cm]{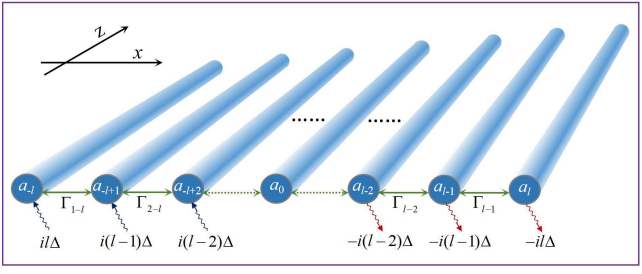}
\caption{A $\mathcal{PT}$-symmetric array of $2l+1$ single-mode waveguides arranged (extending) along the $x$-axis ($z$ axis). Each mode $a_{n}$ exhibits an imaginary propagation constant $-in\Delta$ indicating balanced gain ($n<0$) and loss ($n>0$) for $\Delta>0$ while two real coupling constants $\Gamma_{n}$ and $\Gamma_{n-1}$ ($\Gamma_{n+1}$ and $\Gamma_{n}$) with its neighbors $a_{n+1}$ and $a_{n-1}$, respectively, for $n>0$ ($n<0$).} \label{fig1}
\end{figure}

In this section, we start by considering in Fig.~\ref{fig1} an array consisting of $N=2l+1$ single-mode waveguides labeled by $a_{n}$ for $n\in\{-l,l\}$. 
Our goal is to examine nontrivial light transfer behaviors based on $N$th-order EPs by constructing a $\mathcal{PT}$-symmetric mode-coupling matrix for this $N$-mode waveguide array with underlying $SU(2)$ symmetry. 
Generally speaking, $N$th-order EPs are viable to be observed via a fine tuning of propagation and coupling constants, though may be easily split into lower-order
EPs even in case of a slightly broken $N$th-order $SU(2)$ symmetry.
In principle, two $N$th-order EPs will be found at $\alpha=\pm
\beta$ for an $N\times N$ non-Hermitian matrix $H_{non}=\alpha
S_{i}+i\beta S_{j}$ ($i=x$ and $j=z$ or vice versa), being $S_{i}$
($S_{j}$) the Cartesian operator along the $i$-axis ($j$-axis) of a
spin-$l$ system~\cite{Haml2}. Such non-Hermitian matrices have been
adopted to describe various quantum or classical systems, including
the Bose-Hubbard model~\cite{Haml1},
 ultracold Bose gases~\cite{394}, and waveguide
arrays~\cite{hEPsen3}.

With above considerations, we construct the $\mathcal{PT}$-symmetric mode-coupling matrix 
\begin{widetext}
\begin{equation}
        H=2\Gamma S_{x}+i\Delta S_{z}=
        \left(
        \begin{array}{ccccccc}
            i{l}\Delta & \Gamma_{1-l} & 0 & \cdots  & 0 & 0 & 0 \\
            \Gamma_{1-l} & -i(1-l)\Delta & \Gamma_{2-l} & \cdots   & 0 & 0 & 0 \\
            0 & \Gamma_{2-l} &  -i(2-l)\Delta  & \ddots  &  \ddots &  0 & 0  \\
            \vdots & \vdots & \ddots  & \ddots &  \ddots & \vdots & \vdots \\
            0& 0  & \ddots  &  \ddots & -i(l-2)\Delta & \Gamma_{l-2} & 0 \\
            0 & 0 & 0  & \cdots & \Gamma_{l-2} & -i(l-1)\Delta & \Gamma_{l-1} \\
            0 & 0 &  0 & \cdots & 0 &  \Gamma_{l-1} & -i{l}\Delta  \\
        \end{array}
        \right),\label{Eqn:HPT}
\end{equation}
\end{widetext}
with imaginary diagonal elements $H_{n,n}=-in\Delta$ for
$n\in\{-l,l\}$ and real sub-diagonal elements
$H_{n\pm1,n}=H_{n,n\pm1}=\Gamma_{n}$ restricted by
$\Gamma_{n}=\Gamma\sqrt{(l\pm n+1)(l\mp n)}$ for
$n\in\pm\{0,l-1\}$. With respect to the waveguide array as
illustrated in Fig.~\ref{fig1}~\cite{nonHopt2}, imaginary diagonal
elements refer to gain or loss constants adjusted, \textit{e.g.} by
a temperature gradient~\cite{wavT1,wavT2}, while real sub-diagonal
elements refer to adjacent couplings adjusted, \textit{e.g.} by the
waveguide separations~\cite{wavS1,wavS2}. To be more specific, the
waveguide array described by Eq.~(\ref{Eqn:HPT}) exhibits balanced
gain and loss ($H_{-n,-n}=-H_{n,n}=in\Delta$) in the presence of a
neutral ($H_{0,0}=0$) central waveguide. 
On this basis, the mode-coupling matrix in Eq.~(\ref{Eqn:HPT}) is found to be exactly $\mathcal{PT}$ symmetric because it satisfies the commutation relation $[\mathcal{PT},H]=0$, where the parity operator $\mathcal{P}$ indicates spatial reflection and is described by the standard involutory permutation matrix
\begin{equation}
    \mathcal{P}=\left(
    \renewcommand\arraystretch{1}
    \begin{array}{ccccc}
        0 & 0 & \cdots & 0 & 1 \\
        0 & 0 & \cdots & 1 & 0 \\
        \vdots & \vdots & \ddots & \vdots & \vdots \\
        0 & 1 & \cdots & 0 & 0 \\
        1 & 0 & \cdots & 0 & 0
    \end{array}\right)\label{Eqn:parity},
\end{equation}
while the time reversal operator $\mathcal{T}$ indicates complex
conjugation with $H^{\dagger}=\mathcal{T}H^{T}\mathcal{T}=\mathcal{T}H\mathcal{T}$.

\begin{figure}[ptb]
\centering\includegraphics[width=8.5 cm]{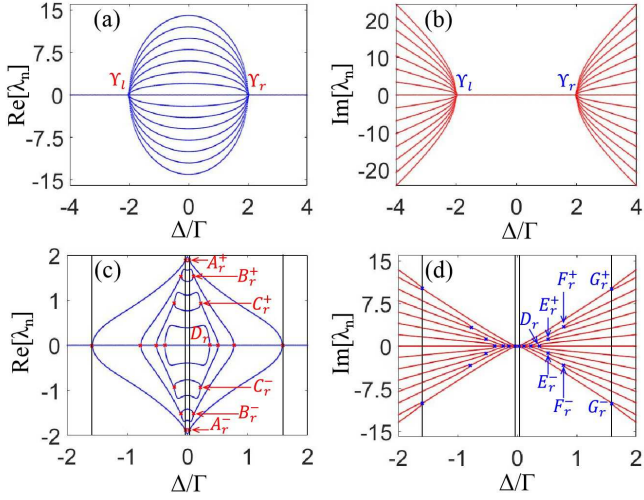} \caption{(a) Real and (b) imaginary parts of $2l+1$ eigenvalues of $\mathcal{PT}$-symmetric mode-coupling matrix $H$ in Eq.~(\ref{Eqn:HPT}) against $\Delta/\Gamma$ with $\Gamma_{n}=\Gamma\sqrt{(l\pm n+1)(l\mp n)}$ for $n\in\pm\{0,l-1\}$ and $l=7$. 
(c) Real and (d) imaginary parts of $2l+1$ eigenvalues of $\mathcal{PT}$-symmetric mode-coupling matrix $H$ in Eq.~(\ref{Eqn:HPT}) against $\Delta/\Gamma$ with $\Gamma_{n}=\Gamma$ for $n\in\pm\{0,l-1\}$ and $l=7$.} \label{fig2}
\end{figure}

We plot real parts in Fig.~\ref{fig2}(a) while imaginary parts in
Fig.~\ref{fig2}(b) of eigenvalues $\lambda_{n}$ of $\mathcal{PT}$-symmetric
mode-coupling matrix $H$ in Eq.~(\ref{Eqn:HPT}) with $l=7$. It is
easy to see that all fifteen eigenvalues coalesce at
$\Delta/\Gamma=\pm2$ with $\lambda_{n}\equiv0$, yielding thus two
highest $15$th-order EPs marked by $\Upsilon_{l}$ and
$\Upsilon_{r}$. It is more important that the $\mathcal{PT}$-symmetric waveguide
array can be engineered on demand to work in the unbroken phase with
purely real eigenvalues in the case of $|\Delta/\Gamma|<2$, while in
the broken phase with purely imaginary eigenvalues in the case of $|\Delta/\Gamma|>2$. To clarify the importance of choosing proper sub-diagonal elements, we then make similar plots in Fig.~\ref{fig2}(c) and Fig.~\ref{fig2}(d) for another $\mathcal{PT}$-symmetric mode-coupling matrix where
$\Gamma_{n}=\Gamma\sqrt{(l\pm n+1)(l\mp n)}$ is replaced by
$\Gamma_{n}=\Gamma$ for $n\in\pm\{0,l-1\}$. 
It is found that each $15$th-order EP degrades into a dozen of $2$nd-order EPs and a single $3$rd-order EP. 
To be more specific, points $\Upsilon_{l}$ and $\Upsilon_{r}$ are split into groups $\{A_{l}^{\pm},B_{l}^{\pm},C_{l}^{\pm},D_{l},E_{l}^{\pm},F_{l}^{\pm},G_{l}^{\pm}\}$ (not shown) and $\{A_{r}^{\pm},B_{r}^{\pm},C_{r}^{\pm},D_{r},E_{r}^{\pm},F_{r}^{\pm},G_{r}^{\pm}\}$ (shown), respectively, in a way of exact mirror symmetry with respect to $\Delta/\Gamma=0$. 
This then results in a very narrow unbroken phase between points $A_{l}^{\pm}$ and $A_{r}^{\pm}$; twelve partially broken phases between points $G_{l}^{\pm}$ and $A_{l}^{\pm}$ as well as between points $A_{r}^{\pm}$ and $G_{r}^{\pm}$; two fully broken phases at the left of points $G_{l}^{\pm}$ or at the right of points $G_{r}^{\pm}$.

As a light beam is incident upon one or more of the $N$ waveguides, its longitudinal propagation along the $z$-axis and transverse transfer 
between adjacent waveguides evolve according to a Schr\"{o}dinger-like equation $i\partial_{z}\psi=H\psi$~\cite{nonHopt1,nonHopt2}. 
Here an $N\times1$ column vector $\psi=(a_{-l},...,a_{0},...,a_{l})^T $ has been introduced to describe the collective excitation of this $\mathcal{PT}$-symmetric waveguide array, being $a_{n}$ an excitation amplitude of the $n$th mode.
Using the PT-symmetric mode-coupling matrix $H$ in Eq.~(\ref{Eqn:HPT}), it is straightforward to expand the Schr\"{o}dinger-like equation into

\begin{align}
\frac{\partial a_{n}}{\partial z}=& \left  \{
    \begin{aligned}
    &-n\Delta a_{n}-i\Gamma_{n}a_{n-1}-i\Gamma_{n+1}a_{n+1} &(n&<0),\\
    &-i\Gamma_{n}(a_{n-1}+a_{n+1}) &(n&=0),\\
    &-n\Delta a_{n}-i\Gamma_{n-1}a_{n-1}-i\Gamma_{n}a_{n+1} &(n&>0),
    \end{aligned}\label{Eqn:motion-PT}
\right.
\end{align}
restricted by the boundary conditions $n\in\{-l,l\}$ and
$\Gamma_{-l-1}=\Gamma_{l+1}=0$.

\begin{figure}[ptb]
\centering\includegraphics[width=8.5 cm]{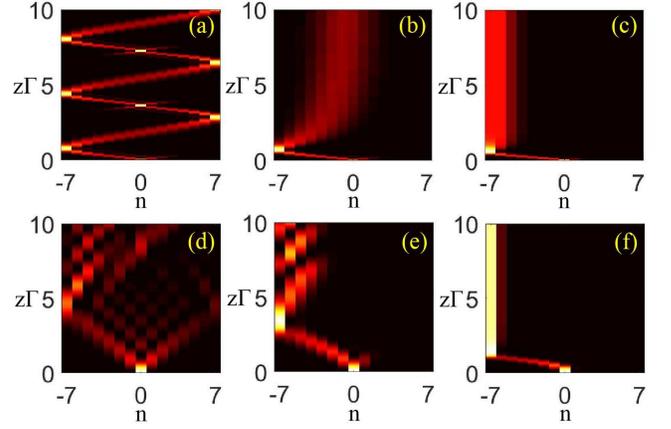}
\caption{Evolutions of mode excitations in a $\mathcal{PT}$-symmetric waveguide
array with $\Gamma_{n}=\Gamma\sqrt{(l\pm n+1)(l\mp n)}$ (a,b,c) or
$\Gamma_{n}=\Gamma$ (d,e,f) for $n\in\pm\{0,l-1\}$ and $l=7$
regarding the same initial excitation $a_{n}(0)=\delta_{n,0}$. The
upper and lower panels are plotted with $\Delta/\Gamma=1$ (a);
$\Delta/\Gamma=2$ (b); $\Delta/\Gamma=4$ (c) and
$\Delta/\Gamma=0.02$ (d); $\Delta/\Gamma=0.494$ (e);
$\Delta/\Gamma=3$ (f), respectively.} \label{fig3}
\end{figure}

We examine with Eq.~(\ref{Eqn:motion-PT}) evolutions of mode excitations characterized by $|a_{n}(z)|^{2}/\sum_{n}|a_{n}(z)|^{2}$ in the $\mathcal{PT}$-symmetric waveguide array for a single-site initial excitation $a_{n}(0)=\delta_{n,0}$. Plotted with $\Delta/\Gamma=1$ in the unbroken phase, Fig.~\ref{fig3}(a) shows that the incident light first transfers unidirectionally toward waveguides of larger gain constants until reaching the left edge, then oscillates symmetrically between the left and right edges with a period $\delta z=2\pi/\sqrt{4\Gamma^{2}-\Delta^{2}}$. 
This is different from the (breathing-like) bidirectional Bloch oscillations observed in corresponding Hermitian waveguide arrays~\cite{wavT1}
because mode-coupling matrix $H$ exhibits real eigenvalues and linearly gradient gain/loss constants. 
In other words, the unidirectional oscillation occurs only if an excitation transfer from the central ($n=0$) waveguide is increasingly evanescent toward the right ($n>0$) waveguides while increasingly amplified toward the left ($n<0$) waveguides in an exactly asymmetric manner. 
Plotted with $\Delta/\Gamma=2$ at one $N$th-order EP $\Upsilon_{r}$, Fig.~\ref{fig3}(b) shows that mode excitations are localized around the central waveguide with no oscillations after a single reflection from the left edge. 
This may be attributed to a strong centripetal force arising from the highest-degree degeneracy of all eigenvalues $\lambda_{n}\equiv0$ and promises a robust directional output of mode excitations from the interface of middle waveguides with loss and gain respectively. Plotted with $\Delta/\Gamma=4$ in the broken phase, Fig.~\ref{fig3}(c) shows that mode excitations suffer neither oscillations nor reflections and are localized near the left edge instead, similar to the skin effect. 
This promises another robust directional output of mode excitations clinging to the left edge and can be understood by considering that complex eigenvalues of mode-coupling matrix $H$ lead to evanescent evolutions in general, leaving only amplified supermodes on waveguides of largest gain~\cite{nonHopt1}. 
Similar light transfer behaviors have been discussed in~\cite{nonHopt1}, but it is worth noting that mode excitations in our work are confined in a much smaller number of waveguides, yielding thus much tighter light transfer behaviors facilitating potential applications, applications, as spin-$l$ matrices $S_x$ and $S_z$ are used to construct mode-coupling matrix $H$. 

\begin{figure}[ptb]
\centering\includegraphics[width=8.5 cm]{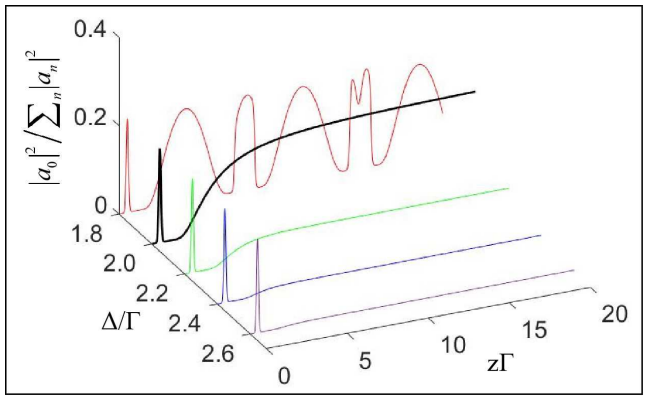}
\caption{Evolutions of central ($n=0$) mode excitation referring to a few typical ratios of $\Delta/\Gamma$ in a $\mathcal{PT}$-symmetric waveguide array with $\Gamma_{n}=\Gamma\sqrt{(l\pm n+1)(l\mp n)}$ for $n\in\pm\{0,l-1\}$ and $l=7$ regarding the initial excitation $a_{n}(0)=\delta_{n,7}$.} \label{fig4}
\end{figure}

Light transfer behaviors in the $\mathcal{PT}$-symmetric waveguide array will become quite different, provided that each $N$th-order EP is split into a series of lower-order EPs as shown in Fig.~\ref{fig2}(c) and Fig.~\ref{fig2}(d). 
This is examined in Fig.~\ref{fig3}(d) for an unbroken phase between $2$nd-order EPs $A_{l}^{\pm}$ and $A_{r}^{\pm}$; 
in Fig.~\ref{fig3}(e) at $2$nd-order EPs $E_{l}^{\pm}$ separating two partially broken phases; 
in Fig.~\ref{fig3}(f) for a fully broken phase at the right of $2$nd-order EPs $G_{r}^{\pm}$. We can see that evolutions of mode excitations vary from an asymmetric (breathing-like) bidirectional oscillation behavior, through an imperfect left-towards localization behavior, to a perfect left-towards localization behavior when $\Delta/\Gamma$ is increased to across one by one the $2$nd and $3$rd-order EPs. 
This nontrivial transition of light transfer behaviors depends critically on the order (degree of degeneracy) and the number of EPs. 
It is worth noting that the center-towards localization behaviors observed in Fig.~\ref{fig3}(b) cannot be observed again in the absence of highest $N$th-order EPs.

Each $N$th-order EP in Fig.~\ref{fig2}(a) and Fig.~\ref{fig2}(b) for the $\mathcal{PT}$-symmetric waveguide array indicates also a sensitive transition with respect to distinct light transfer behaviors. 
This is justified in Fig.~\ref{fig4} by examining $|a_{0}(z)|^{2}/\Sigma_{n}|a_{n}(z)|^{2}$ for a few typical ratios of $\Delta/\Gamma$ and another single-site initial excitation $a_{n}=\delta_{n,7}$. It is easy to see that the central waveguide exhibits a robust mode excitation with neither periodic oscillations nor a fast decay for $\Delta/\Gamma=2$ referring to one $N$th-order EP $\Upsilon_{r}$. 
Adjusting $\Delta/\Gamma$ to enter the unbroken or broken phase beyond this $N$th-order EP, we find instead periodic oscillations or a fast decay of mode excitation in the central waveguide. 
Hence it is viable to attain a robust output of mode excitation from the central waveguide or realize a sensitive switching between distinct light transfer behaviors in the $\mathcal{PT}$-symmetric waveguide array by accurately modulating $\Delta$ or $\Gamma$.

\section{anti-$\mathcal{PT}$-symmetric waveguide array}

In this section, we try to construct an anti-$\mathcal{PT}$-symmetric mode-coupling matrix so as to further investigate highest $N$th-order EPs and nontrivial light transfer behaviors in an alternative array consisting of $N=2l+1$ single-mode waveguides.
This can be done by rotating both $S_{x}$ and $S_{z}$ in the $\mathcal{PT}$-symmetric mode-coupling matrix $H$ around $S_{y}$, yielding thus $S_{x}\to S_{z}$ and $S_{z}\to -S_{x}$. 
The resultant anti-$\mathcal{PT}$-symmetric mode-coupling matrix is

\begin{widetext}

    \begin{equation}
        \widetilde{H}=\Delta S_{z}-2i\Gamma S_{x}=
        \left(
        \begin{array}{ccccccc}
            l\Delta & -i\Gamma_{1-l} & 0 & \cdots  &0  & 0 & 0 \\
            -i\Gamma_{1-l} &(l-1)\Delta & -i\Gamma_{2-l} & \cdots  & 0 & 0 & 0 \\
            0 & -i\Gamma_{2-l} & (l-2)\Delta  &  \ddots &  \ddots &  0 &  0 \\
            \vdots & \vdots & \ddots  & \ddots &  \ddots & \vdots & \vdots \\
            0&  0 & \ddots  & \ddots  & -(l-2)\Delta & -i\Gamma_{l-2} & 0 \\
            0 & 0 & 0  & \cdots & -i\Gamma_{l-2} & -(l-1)\Delta & -i\Gamma_{l-1} \\
            0 & 0 & 0  & \cdots & 0 & -i\Gamma_{l-1} & -{l}\Delta  \\
        \end{array}\label{Eqn:HAPT}
        \right),
    \end{equation}
\end{widetext}
which reserves the underlying $SU(2)$ symmetry on one hand and satisfies the anti-commutation relation $\{\mathcal{PT},\widetilde{H}\}=0$ on the other hand. 
Obviously, the $2l+1$ diagonal elements $\widetilde{H}_{n,n}$ become real so that all waveguide modes $a_{n}$ suffer neither loss nor gain, while the $4l$ sub-diagonal elements $\widetilde{H}_{n,n\pm1}$ become imaginary so that waveguide modes $a_{n}$ and $a_{n\pm1}$ are passively coupled. 
However, passive couplings are usually not realistic, hence one question arises: is it viable to realize mode-coupling matrix $\widetilde{H}$ by modifying the waveguide array as shown in Fig.~\ref{fig1}?

\begin{figure}[ptb]
\centering\includegraphics[width=8.5 cm]{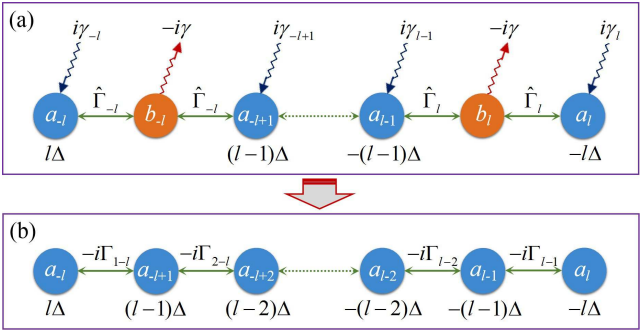} 
\caption{(a) A non-Hermitian waveguide array consisting of $2l+1$ primary modes $a_{n}$ and $2l$ assistant modes $b_{n}$. Each $b_{n}$ exhibits an imaginary propagation constant $-i\gamma$ (loss for $\gamma>0$) and a real coupling constant $\widehat{\Gamma}_{n}$ with two adjacent $a_{n}$, which exhibits instead a complex propagation constant $-n\Delta+i\gamma_{n}$ (gain for $\gamma_{n}>0$). 
(b) An equivalent anti-$\mathcal{PT}$-symmetric waveguide array with imaginary coupling constants $-i\Gamma_{n}$ after eliminating the $2l$ assistant modes with properly designed $\gamma_{n}$ and $\widehat{\Gamma}_{n}$.}
\label{fig5}
\end{figure}

The answer is positive provided we insert an assistant
waveguide described by mode $b_{n}$ between each pair of primary waveguides described by modes $a_{n}$ and $a_{n-1}$ ($a_{n+1}$) for $n>0$ ($n<0$) as shown in Fig.~\ref{fig5}(a)~\cite{loss3}. With respect to this $\widehat{N}=4l+1$ waveguide array, we can write down the Schr\"{o}dinger-like equation $i\partial_{z}\widehat{\psi}=\widehat{H}\widehat{\psi}$ for a $\widehat{N}\times \widehat{N}$ mode-coupling matrix $\widehat{H}$ and a $\widehat{N}\times1$ column vector $\widehat{\psi}$ and expand it into the following equations of motion

\begin{subequations}
\begin{align}
\frac{\partial a_{n}}{\partial z}=& \left  \{
    \begin{aligned}
    &(in\Delta+\gamma_{n}) a_{n}-i\widehat{\Gamma}_{n-1}b_{n-1}-i\widehat{\Gamma}_{n}b_{n} &(n&<0),\\
    &\gamma_{n}a_{n}-i\widehat{\Gamma}_{n-1}b_{n-1}-i\widehat{\Gamma}_{n+1}b_{n+1} &(n&=0),\\
    &(in\Delta+\gamma_{n}) a_{n}-i\widehat{\Gamma}_{n}b_{n}-i\widehat{\Gamma}_{n+1}b_{n+1} &(n&>0);
    \end{aligned}\label{Eqn:motion-primary}
\right.\\
\frac{\partial b_{n}}{\partial z}=& \left  \{
    \begin{aligned}
    &-\gamma b_{n}-i\widehat{\Gamma}_{n}(a_{n}+a_{n+1}) &(n&<0),\\
    &-\gamma b_{n}-i\widehat{\Gamma}_{n}(a_{n-1}+a_{n}) &(n&>0),
    \end{aligned}\label{Eqn:motion-assistant}
\right.\\\notag
\end{align}
\end{subequations}
where $\gamma_{n}$ describes different gain constants of primary modes $a_{n}$ while $\gamma$ represents the common decay rates of assistant modes $b_{n}$. 
We have also introduced $\widehat{\Gamma}_{n}$ to denote different coupling constants between modes $b_{n}$ and $a_{n}$ as well as $b_{n}$ and $a_{n+1}$ ($a_{n-1}$) for $n<0$ ($n>0$). In the limit of $\gamma\gg \widehat{\Gamma}_{n}$, it is viable to attain with Eq.~(\ref{Eqn:motion-assistant})

\begin{align}
b_{n}=& \left  \{
    \begin{aligned}
    &-i\widehat{\Gamma}_{n}(a_{n}+a_{n+1})/\gamma &(n&<0),\\
    &-i\widehat{\Gamma}_{n}(a_{n-1}+a_{n})/\gamma &(n&>0),
    \end{aligned}\label{Eqn:steady-assistant}
\right.\\\notag
\end{align}
by directly setting $\partial_{z} b_{n}=0$ with the approximation of an adiabatic evolution~\cite{loss3,loss5}.

Substituting Eq.~(\ref{Eqn:steady-assistant}) back into Eq.~(\ref{Eqn:motion-primary}), we can further attain
\begin{widetext}
\begin{align}
\frac{\partial a_{n}}{\partial z}=& \left  \{
    \begin{aligned}
    &[in\Delta+\gamma_{n}-(\widehat{\Gamma}_{n-1}^{2}+\widehat{\Gamma}_{n}^{2})/\gamma]a_{n}
    -(\widehat{\Gamma}_{n-1}^{2}/\gamma)a_{n-1}-(\widehat{\Gamma}_{n}^{2}/\gamma)a_{n+1} &(n&<0),\\
    &[\gamma_{n}-(\widehat{\Gamma}_{n-1}^{2}+\widehat{\Gamma}_{n+1}^{2})/\gamma]a_{n}
    -(\widehat{\Gamma}_{n-1}^{2}/\gamma)a_{n-1}-(\widehat{\Gamma}_{n+1}^{2}/\gamma)a_{n+1} &(n&=0),\\
    &[in\Delta+\gamma_{n}-(\widehat{\Gamma}_{n}^{2}+\widehat{\Gamma}_{n+1}^{2})/\gamma]a_{n}
    -(\widehat{\Gamma}_{n}^{2}/\gamma)a_{n-1}-(\widehat{\Gamma}_{n+1}^{2}/\gamma)a_{n+1}
    &(n&>0),
    \end{aligned}\label{Eqn:motion-reduced-1st}
\right.
\end{align}
\end{widetext}
by eliminating the $2l$ assistant modes $b_{n}$. 
Eq.~(\ref{Eqn:motion-reduced-1st}) finally reduces into

\begin{align}
\frac{\partial a_{n}}{\partial z}=& \left  \{
    \begin{aligned}
    &i n\Delta a_{n}-\Gamma_{n}a_{n-1}-\Gamma_{n+1}a_{n+1} &(n&<0),\\
    &-\Gamma_{n}(a_{n-1}+a_{n+1}) &(n&=0),\\
    &i n\Delta a_{n}-\Gamma_{n-1}a_{n-1}-\Gamma_{n}a_{n+1} &(n&>0),
    \end{aligned}\label{Eqn:motion-reduced-2nd}
\right.
\end{align}
when we take ($i$) $\gamma_{n}=(\widehat{\Gamma}_{n-1}^{2}+\widehat{\Gamma}_{n}^{2})/\gamma$ for $n<0$, $\gamma_{0}=(\widehat{\Gamma}_{-1}^{2}+\widehat{\Gamma}_{1}^{2})/\gamma$ for $n=0$, and $\gamma_{n}=(\widehat{\Gamma}_{n}^{2}+\widehat{\Gamma}_{n+1}^{2})/\gamma$ for $n>0$; 
($ii$) $\widehat{\Gamma}_{n}^{2}/\gamma=\Gamma_{n+1}$ for $n<0$ and $\widehat{\Gamma}_{n}^{2}/\gamma=\Gamma_{n-1}$ for $n>0$. 
That means, we have $\widehat{\Gamma}_{-n}^{2}=\widehat{\Gamma}_{n}^{2}$ and $\gamma_{-n}=\gamma_{n}$ for $n\in\pm\{1,l\}$ restricted again by $\Gamma_{n}=\Gamma\sqrt{(l\pm n+1)(l\mp n)}$ for $n\in\pm\{0,l-1\}$ as the array of $\widehat{N}$ coupled waveguides in Fig.~\ref{fig5}(a) reduces to the array of $N$ waveguides in Fig.~\ref{fig5}(b). 
Eq.~(\ref{Eqn:motion-reduced-2nd}) is what we can attain by inserting the $N\times1$ column vector $\psi=(a_{-l},...,a_{0},...,a_{l})^{T}$ and the $N\times N$ mode-coupling matrix $\widetilde{H}$ into the Schr\"{o}dinger-like equation $i\partial_{z}\psi=\widetilde{H}\psi$, justifying the feasibility for realizing effectively neutral waveguide arrays with passive couplings.

\begin{figure}[ptb]
\centering\includegraphics[width=8.5 cm]{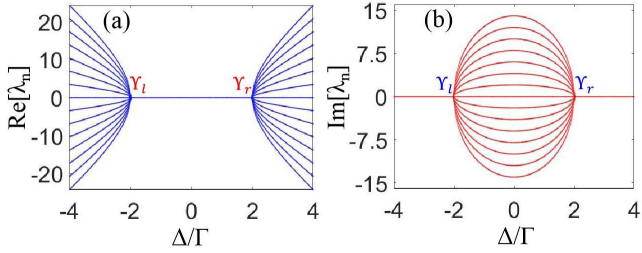} 
\caption{(a) Real and (b) imaginary parts of $2l+1$ eigenvalues of anti-$\mathcal{PT}$-symmetric mode-coupling matrix $\widetilde{H}$ in Eq.~(\ref{Eqn:HAPT}) against $\Delta/\Gamma$ with $\Gamma_{n}=\Gamma\sqrt{(l\pm n+1)(l\mp n)}$ for $n\in\pm\{0,l-1\}$ and $l=7$.} \label{fig6}
\end{figure}

We plot in Fig.~\ref{fig6}(a) and Fig.~\ref{fig6}(b), respectively, real and imaginary parts of eigenvalues $\lambda_{n}$ of the anti-$\mathcal{PT}$-symmetric matrix $\widetilde{H}$ in Eq.~(\ref{Eqn:HAPT}) with $l=7$. In comparison to that observed in Fig.~\ref{fig2}(a) and
Fig.~\ref{fig2}(b), the common feature is that all fifteen eigenvalues coalesce again at $\Delta/\Gamma=\pm2$ with $\lambda_{n}\equiv0$ to yield two highest $15$th-order EPs $\Upsilon_{l}$ and $\Upsilon_{r}$; the difference lies in that the unbroken and broken phases refer instead to purely imaginary and purely real eigenvalues, respectively~\cite{loss5}. 
To be more specific, the unbroken (broken) phase is attained with a smaller (larger) absolute ratio of imaginary propagation constant $-in\Delta$ and real coupling constant $\Gamma_{n}$ for $H$ in Eq.~(\ref{Eqn:HPT}), while with a smaller (larger) absolute ratio of real propagation constant $-n\Delta$ and imaginary coupling constant $-i\Gamma_{n}$ for $\widetilde{H}$ in Eq.~(\ref{Eqn:HAPT}). 
Therefore, distinct light transfer behaviors are expected to occur in the $\mathcal{PT}$-symmetric and anti-$\mathcal{PT}$-symmetric waveguide arrays even if they are engineered to work with the same parameters as discussed below.

\begin{figure}[ptb]
\centering\includegraphics[width=8.5 cm]{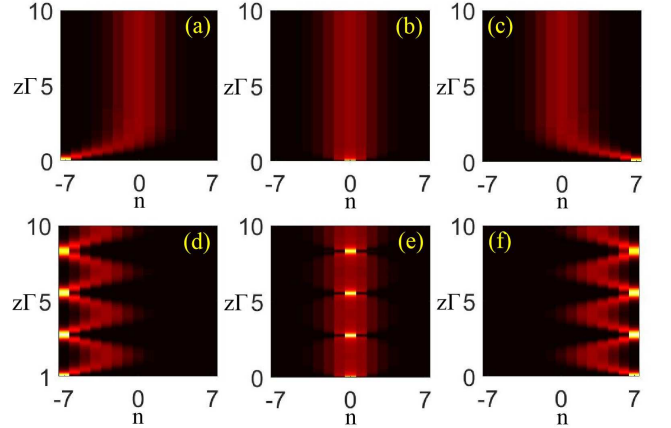}
\caption{Evolutions of mode excitations in an anti-$\mathcal{PT}$-symmetric waveguide array with $\Gamma_{n}=\Gamma\sqrt{(l\pm n+1)(l\mp n)}$ for $n\in\pm\{0,l-1\}$ and $l=7$ regarding different initial excitations $a_{n}(0)=\delta_{n,-7}$ (a,d); $a_{n}(0)=\delta_{n,0}$ (b,e); $a_{n}(0)=\delta_{n,7}$ (c,f). 
The upper and lower panels are plotted with $\Delta/\Gamma=1$ (a,b,c) and $\Delta/\Gamma=3$ (d,e,f), respectively.} \label{fig7}
\end{figure}

We then examine with Eq.~(\ref{Eqn:motion-reduced-2nd}) evolutions of mode excitations characterized by $|a_{n}(z)|^{2}/\sum_{n}|a_{n}(z)|^{2}$ in the anti-$\mathcal{PT}$-symmetric waveguide array for different single-site initial excitations $a_{n}(0)=\delta_{n,-7}$, $a_{n}(0)=\delta_{n,0}$, and $a_{n}(0)=\delta_{n,7}$ as shown in Fig.~\ref{fig7}. 
In the unbroken phase with $\Delta/\Gamma=1$, we can see from Figs.~\ref{fig7}(a-c) that mode excitations always converge to and finally leave from a few middle waveguides without exhibiting oscillations and independent of initial single-site excitations.
Such a robust center-towards localization behavior is clearly different from the left-edge-towards localization behavior observed in the broken phase of an $\mathcal{PT}$-symmetric waveguide array as shown in Fig.~\ref{fig3}(c). 
The underlying physics for this center-towards localization behavior should be that all adjacent waveguides are passively coupled with the gradually reduced imaginary coupling constants $-i\Gamma_{n}=-i\Gamma_{-n}$ from both edges to the center. 
In the broken phase with
$\Delta/\Gamma=3$, we can see from Figs.~\ref{fig7}(d-f) that mode excitations oscillate with an identical fixed period determined by the ratio $\Delta/\Gamma$ but of different centers identical to origins of single-site excitations. This more localized
origin-centered oscillation may be attributed to a larger ratio
$\Delta/\Gamma=3$ and is also very different from that observed in the unbroken phase of an $\mathcal{PT}$-symmetric waveguide array as shown in Fig.~\ref{fig3}(a).

\begin{figure}[ptb]
\centering\includegraphics[width=8.5 cm]{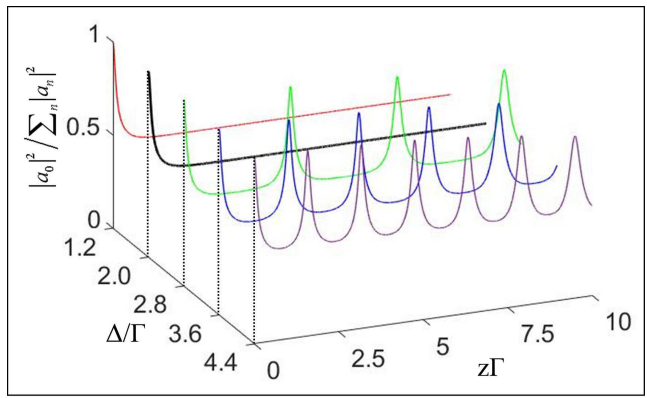}
\caption{Evolutions of central ($n=0$) mode excitations referring to
different ratios of $\Delta/\Gamma$ in an anti-$\mathcal{PT}$-symmetric
waveguide array with $\Gamma_{n}=\Gamma\sqrt{(l\pm n+1)(l\mp n)}$
for $n\in\pm\{0,l-1\}$ and $l=7$ regarding the initial excitation
$a_{n}(0)=\delta_{n,0}$.}\label{fig8}
\end{figure}

To gain a more intuitive insight into the transition between
different light transfer behaviors, we focus on the central ($n=0$)
waveguide by examining $|a_{0}(z)|^{2}/\Sigma_{n}|a_{n}(z)|^{2}$ in
Fig.~\ref{fig8} for different values of $\Delta/\Gamma$ while a
fixed initial excitation $a_{n}=\delta_{n,0}$. As the
anti-$\mathcal{PT}$-symmetric waveguide array works at and below its $N$th-order
EP $\Upsilon_{r}$, consistent with that observed in
Fig.~\ref{fig7}(b) but different from that observed in
Fig.~\ref{fig4}, the central mode excitation quickly decays to a steady value $\sim 0.5$ independent of the ratio $\Delta/\Gamma$. 
As the anti-$\mathcal{PT}$-symmetric waveguide array works above the $N$th-order EP $\Upsilon_{r}$, consistent with that observed in Fig.~\ref{fig7}(e) but different from that observed in Fig.~\ref{fig4}, the central mode excitation starts to oscillate with a period $\delta z=2\pi/\sqrt{\Delta^{2}-4\Gamma^{2}}$ tunable in an arbitrarily large range (unattainable yet in the $\mathcal{PT}$-symmetric waveguide array) since $\Delta/2\Gamma$ may approach either unit or infinity. 
Hence, it is viable to attain a controlled output of mode excitation from the central waveguide by adjusting the ratio $\Delta/\Gamma$, which may serve as an efficient scheme of high-precision
measurement on relevant parameters determining $\Delta$ or $\Gamma$.

\section{Conclusions}

In summary, we have studied nontrivial light transfer dynamics by
working at and beyond one of two $N$th-order EPs, referring to
highest-degree degeneracies of all $N$ eigenvalues of relevant
mode-coupling matrices, in two non-Hermitian waveguide arrays
exhibiting a $\mathcal{PT}$ or an anti-$\mathcal{PT}$ symmetry. For the $\mathcal{PT}$-symmetric
waveguide array with balanced gain and loss, light transfer dynamics
is viable to switch between a unidirectional oscillation behavior in
the unbroken phases, through a center-towards localization behavior
at one $N$th-order EP, to an edge-towards localization behavior in
the broken $\mathcal{PT}$ phases. For the anti-$\mathcal{PT}$-symmetric waveguide array with
passive couplings, light transfer dynamics is very different again
in the broken and unbroken phases, hence supporting the transition
between an origin-centered oscillation behavior and a center-towards
localization behavior across one $N$th-order EP. These drastic
transitions of light transfer behaviors are expected to facilitate
advanced photonic applications like sensitive optical switching and
high-precision measurement, while providing also new perspectives
for further theoretical studies.

\section*{Acknowledgments}

This work is supported by National Natural Science Foundation of
China (NO.~11704064 and No.~12074061), Fundamental Research Funds
for the Central Universities (NO.~2412019FZ045), and Science
Foundation of the Education Department of Jilin Province
(NO.~JJKH20211279KJ).


\end{document}